\documentclass[fleqn,10pt]{olplainarticle}

\usepackage[english]{babel}
\usepackage{xcolor, soul, colorspace}
\usepackage{enumitem, supertabular}
\usepackage{url}

\definecolor{coolgray1c}{RGB}{217,217,214}
\definecolor{lovebird}{RGB}{193,219,60}
\definecolor{illuminating}{RGB}{245,223,77}
\definecolor{PMS1585C}{RGB}{255,106,20}
\definecolor{PMS2905C}{RGB}{158,201,227}

\title{Finding Structure in Silence: The Role of Pauses in Aligning Speaker Expectations}

\author{Maja Linke$^1$}
\author{Michael Ramscar$^2$}
\affil{
$^1$Max Planck Institute for Human
Cognitive and Brain Sciences, Leipzig, Germany \\
$^2$Linguistics Department\unskip, University of T{\"{u}}bingen\unskip, T{\"{u}}bingen\unskip, Germany
}

\keywords{signal/noise discrimination; memorylessness; information structure; speaker alignment; distributed learning}

\begin{abstract}

The intelligibility of speech relies on the ability of interlocutors to dynamically align their expectations about the rates at which informative changes in signals occur. Exactly how this is achieved remains an open question. We propose that speaker alignment is supported by the statistical structure of spoken signals and show how pauses offer a time-invariant template for structuring speech sequences. Consistent with this, we show that pause distributions in conversational English and Korean provide a memoryless information source. We describe how this can facilitate both the initial structuring and maintenance of predictability in spoken signals over time, and show how the properties of this signal change predictably with speaker experience. These results indicate that pauses provide a structuring signal that interacts with the morphological and rhythmical structure of languages, allowing speakers at all stages of lifespan development to distinguish signal from noise and maintain mutual predictability in time.

\end{abstract}

\begin{document}

\flushbottom
\maketitle
\section{The problem of alignment in speech}

Understanding how speakers align their expectations about the occurrence of acoustic events in speech signals is central to explaining the human capacity for vocal communication. However, the challenges this involves are easily obscured by intuitions informed by reading and writing. In contrast to reading and writing, which are self-paced and explicitly taught, speech is highly dependent on timing and is usually learned implicitly through exposure to continuous signals. Meanwhile, although the orderly and discrete way in which letters, words, and phrases appear in texts can misleadingly imply that speech is based on similar, corresponding alphabets of spoken gestures and inventories of discrete forms, spontaneous speech signals are produced by dynamic kinematic processes that guarantee a considerable amount of noise and deviation in the way speech sounds and sequences are produced, such that many of the 'acoustic segments' listeners 'extract' from speech signals are not actually present in the physical stimulus \citep{port2005against, ernestus2002recognition}. The differences between the measurable attributes of physical speech signals and their appearance to receivers pose a deep puzzle: while it is common, and at some level perhaps necessary, to talk about articulation in speech production \citep{goldman1961significance,miller1984articulation} and segmentation in speech comprehension \citep{liberman1967perception,cutler1999comprehending}, the array of findings showing that that much of what listeners 'segment' was never 'articulated' in the first place, while much of what is actually articulated is never segmented \citep{warren1970perceptual, samuel2020psycholinguists}, raise a question that has yet to be resolved: how do speakers bring order to --- and make sense of --- the apparent chaos?

In what follows, we approach this question as a probabilistic puzzle rooted in information theory, seeking to resolve an apparent contradiction between the formal definition of information, and the specific kind of  discrete, periodic structure it requires, on one hand \citep[see][p. 17ff]{shannon1948mathematical}, and the apparent absence of this structure in human communicative signals on the other \citep[see][for discussion]{port2005against, ramscar2019source, linke2020grammar}. Applying the notion of information to speech is consistent with current models that treat spoken communication as a probabilistic process that relies on structured regularities in speech signals \citep[e.g.][]{bell2003effects, aylett2004smooth, bell2009predictability, tily2009syntactic, wedel2013high, seyfarth2014word, seyfarth2016dynamic, wedel2018phonetic, hall2018role, priva2018interdependence}, such that speech perception is viewed as a probabilistic process in which hearers attempt to infer the most likely intended message from a noisy acoustic signal \citep{clayards2008perception, mitterer2009processing, kleinschmidt2015robust}.

However, this probabilistic conception of communication serves to highlight a critical difference between speech codes (and natural languages) and other information-theoretic codes, namely that the former have to be learned. This is a problem because of the highly skewed nature of lexical distributions (which, when aggregated, approach power laws) \citep{estoup1916gammes, zipf1949human}: the long-tailed nature of these distributions results in low-frequency words being irregularly distributed across samples, and guarantee that any individual speaker will only ever experience an incomplete sample of code \citep{ramscar2014myth,ramscar2019source}. This, in turn, guarantees that each speaker's individual experience of codes -- and hence their internal models of them -- will inevitably be unique, such that any individual's experience will necessarily differ from any group average.

The variability in individual speakers' models of expectations raises a problem for the whole idea of statistical regularities in speech signals. Formally, what makes a signal a signal in information theory  is that it comprises a set of structured probabilistic events that are defined by a shared code. Anything else is noise. If speech signals are signals in the same sense, then this raises the question of how speakers ever manage to learn the shared set of probabilities that structures them. In other words, how do speakers ever manage to learn to align their individual expectations about the structure of spoken signals, since this would appear to be necessary to make statistical regularities in the temporal and acoustic properties of the signal predictable and informative in the first place?

\subsection{How do language users organize their expectations?}

Part of the answer to this question lies in findings showing that speech signals that are easily understood in context often become unintelligible when they are presented in isolation \citep{pollack1963intelligibility, bard1983unintelligibility, ernestus2002recognition}, which suggests that speech codes make considerable use of context. However, these findings also raise another question: what, exactly, is context? One recent, successful approach to answering this question has been to define contexts in terms of word collocations in text, using  this information to operationalize the `semantic similarities' between words as similarities between the collocational contexts in which they occur \citep{harris1954distributional, lund1996producing, landauer1997solution, sahlgren2008distributional, mcdonald2001testing, gleitman2002verbs}. Similarly, studies of the empirical structure of word frequency distributions further support the idea that communicative context, in a broader sense, is set by systematic patterns of co-variation between informative events, such as for example speech phrases, words, syllables, and sounds. Analyses of linguistic distributions at multiple levels of description (speech segments, words, and syntactic phrases) show that learnable distributional regularities organize speech sequences in clusters of informative contrasts that provide sites of predictable variation, the information \citep{ramscar2019source, ramscar2021children, linke2020grammar}.

These results suggest that lexical regularities are a source of information for increasing the predictability of signal sequences, indicating that lexical and grammatical forms might support the alignment process \citep{blevins2016morphology, ramscar2018morphological}. However, this suggestion simply serves to underline the problem outlined above: not only does the production and recognition of words rely on context, but in fact, it is often the case that many of the 'acoustic segments' presupposed by the acoustic model (dictionary form, that models a word produced in isolation) are not actually present in the speech signal, purple{but rather} can only be inferred in context \citep{port2005against, ernestus2002recognition}. Moreover, this 'context' is not merely defined by the presence or absence of articulated parts of the signal, but also by the timing of their articulation \citep{dilley2010altering, morrill2014distal, baese2019not, lamekina2022entrainment}. Given that the identification of individual aspects of speech signals such as 'lexical regularities' relies on the alignment of speaker expectations, it follows that these regularities themselves can only be noticed when learners have managed to extract some structure from signals in the first place. This then raises further questions: is there a consistent, time-invariant source of information to support the alignment process? That is, if learners are to be able to initially structure their expectations about speech codes to extract/infer information from signals, and if these systematic patterns of expectation are to be maintained across speakers regardless of their individual experience, it follows that some objective source of information must be available in order to facilitate this. What is it?

One possible source of information that would clearly seem to sidestep these problems is silence, or more specifically, pauses: the regular manifestations of silence in the speech signal. It has long been suggested that coordination between speakers in turn-taking is achieved by a process known as production rate entrainment (which involves the alignment of the rates at which auditory events are transmitted and processed). Variations in pause duration have been argued to play a crucial role in this process \citep{wilson2005oscillator}. Further, at least at the level of conversational turns, silent intervals  across languages converge on remarkably constant averages \citep{stivers2009universals, weilhammer2003durational}, which indicates that any variability in silent interval duration is relatively invariant in time. 

In what follows, we examine whether these apparent structural regularities extend beyond turn-taking. Is temporal variation in speech pauses systematic, and can it play a role in aligning the expectations of speakers and listeners at the level of speech production and recognition as well?

Several empirical characteristics of speech pauses indicate that they could play an important role in the alignment process. First, silence -- or more specifically the absence of articulated signal -- is something that is clearly and 'objectively' present in the speech signal. Second, sensitivity to silence does not seem to rely on experience. Instead, this perceptual contrast appears to emerge prenatally and seems to structure other, more primitive forms of vocalization \citep{mampe2009newborns, wermke2021melody} and serve as a first cue to speech segmentation \citep{mannel2009pauses, seidl2008developmental, holzgrefe2018infants}. This is in contrast to sensitivity to other more complex prosodic patterns and speech sounds, which vary across languages and require more time and experience to learn. Temporal variation in the articulated and the silent parts of the signal also appears to follow distinct patterns across the lifespan. Experience leads to more individual variability in articulation rates, whereas individual variabilities in pause production decrease. In particular, as age and experience increase, the average temporal resolution of the vocal signal (speech rate) increases and becomes more variable between individual speakers and speech contexts \citep{quene2005modeling, jacewicz2010between, hazan2014emergence, tucker2021speech}. By contrast, the duration and variability of speech pauses in conversation is surprisingly stable across the lifespan \citep{redford2013comparative, neuberger2013temporal, bona2011disfluencies, bona2014temporal, hazan2014emergence}.

The absence of effects of experience on pause durations might seem surprising, however, these findings begin to make sense when considered against the backdrop of the mechanisms involved in their processing. In particular, the differential contribution of domain-general and domain-specific mechanisms involved in speech perception and speech production suggests that speech production and speech perception will develop differently over time \citep[and this actually seems to be the case, see][]{campbell2016robust, campbell2018language}. In the brain, the discrimination of "general acoustic events" (including non-events) appears to rely on specialized auditory mechanisms which are subject to lifelong adaptation \citep{yan2003canadian}, whereas the sequential coordination of the motor processes involved in articulation (and thus articulated durations) involves more general timing circuits that subserve a variety of other capacities that involve orientation in space and time \citep{marien2001lateralized, ackermann2007contribution}. Depending on prior experience, some of these latter processes appear to be 'fixed' such that they function independently of the requirements of the specific cognitive task \citep{krakauer2006generalization, wong2017reaction}. Simultaneously, it seems that in both speech perception and production, articulatory events and the variances in their execution are highly context-specific, such that the uncertainties of the contexts and the variances in the way articulations are performed within them constitute a critical part of speaker expectation \citep{tremblay2008specificity, maslowski2019listeners}. In other words, both the execution of these highly automatized motor behaviors and their perception tend not to generalize across modalities/contexts. It follows as a consequence of these considerations that the constraints on timing that articulation imposes must in turn place boundaries on the variability of the rates at which the informative changes in speech can be produced. Accordingly, given the systematic limitations on people's ability to reliably judge noticeable differences with respect to segment duration \citep{friberg1995time, quene2007just}, it seems that when it comes to perception, the way in which the speech signal is interpreted across hearers must be similarly bounded. All of the above points to a likely source of misalignment in speaker/hearers: the diverging experience-driven changes in the temporal distribution of informative auditory events within the limits imposed by the converging temporal resolution of articulations.

Speech directed at children and preverbal infants is more informative in its temporal structure and contains less information in its spectral variation than speech directed at adults \citep{fernald1989intonation, bard1983unintelligibility}. Notably, the extent to which distinct dimensions of prosodic variation (i.e., pauses, pitch, and phrase final lengthening) inform infants' perception and segmentation appears to vary with infants' age and native language \citep{mannel2009pauses, mannel2011intonational, sundara2011rhythmic, mannel2013role, skoruppa2013development, sundara2015perception}. With experience infants' attention gradually shifts away from prosody toward more complex structural regularities in speech signals. However, pauses seem to remain a critical cue to structure throughout childhood and across the adult lifespan \citep{mueller2008role, pena2002signal, mannel2016neural}. Whereas pause durations and the syntactic structures in which they occur do not seem to change substantially across adulthood, the way older speakers realize words does appear to become increasingly context- and speaker-specific \citep{lieberman1989sentence}. Critically, older speakers' vowel realization, both in terms of duration and the relationship between duration and formant dispersion, becomes more variable in connected speech \citep{munson2011phonological, fletcher2015relationship, gahl2019twenty}. Notably, these effects do not transfer to experiments where words are produced in isolation \citep[see][]{watson2007comparison}. This is important because vowels account for the largest part of the variability in word and syllable durations in English, and these developments seem to suggest that the temporal resolution of conversational speech signals at the word and syllable levels change continuously across the lifespan. Conversely, higher-level regularities such as syntactic phrasing and phrase-level segmentation remain relatively stable.

This general idea -- that learning inevitably changes individual models of the world -- is further supported by studies of memory performance in healthy older speakers \citep{ramscar2013learning, ramscar2017mismeasurement}. The processes revealed in these studies seem to guarantee that aspects of the signal that are informative in early communicative experience (e.g., timing, prosody, phrase) will become increasingly uninformative (i.e., stable) as experience grows, and sensitivity to the information provided by fine-grained articulatory variation increases.  These differences lie at the heart of the problem of alignment in speech. The idea that speech signals are informative is largely uncontroversial. However,  formal measures of information ultimately rely on the existence of objective 'events', predictable changes in the temporal and acoustic properties of the signal that are used consistently by all speakers. As we have outlined above, the rates at which information is extracted from the signal will change with speaker experience. Taken together, the considerations above point to an obvious problem when it comes to speech information: because what counts as an informative speech event will vary with experience, most articulated parts of speech signals do not seem compatible with any formal notion of information.

These problems appear to stand in opposition to the idea that the information provided by phonemes, morphemes, or words can play a central role in speaker alignment. Instead, they suggest that speakers must first achieve alignment in relation to some objective rate at which events occur in speech, in order for them to adapt their expectations and hence be able to extract the informative events (and the systematic relations between them) from the signal \citep{edeline1993receptive, morrill2014distal, finn2015segmentation, xie2017more}. Given that articulations vary both physically and temporally, and that human vocal communication (and learning) rely on speakers' ability to anticipate these events in time \citep{patel2006musical, patel2021vocal}, it seems that an aspect of the signal that is stable in at least one of these dimensions is required for these processes to occur. Which brings us back to pauses: can they provide this stable source of information in the speech signal?

As we noted above, a basic prerequisite of an alignment signal is that the information it provides must be invariant across speakers and available to learners of all levels of experience. Formally, in a community of speakers whose shared experience of a class of events differs, alignment will only be possible if the rate at which the events from this class reoccur is somehow independent of the degree to which speakers have sampled those events (i.e., if the results of sampling are somehow independent of experience). Yet all and any of the articulated parts of the speech signal appear to be anything but invariant. Accordingly, it follows that if pauses, which by definition are not articulated, actually do serve as an alignment signal, then we should expect them to be distributed in signals in such a way that the information communicated by them will be independent of speaker experience (within some minimum bound of experience \citep{shannon1948mathematical}). Critically, given that the properties of distributions are determined by both how we define their constituents (what is measured) and their boundaries (how the observation space is limited) it is important that we be clear about what we mean by 'speech pauses' before we proceed with our analysis. 

\section{What are pauses and what do they do?}
\label{pauses}

Not every silence associated with speaking is a pause. Moreover, it is clear that even the events we might theoretically term "pauses" do not form a single coherent class. Previous work has shown that the durations of silent intervals in speech follow a log-normal distribution, which in turn appears to be the product of aggregating (at least) two distinct components: \textbf{gaps}, silent intervals between speaker turns, and \textbf{pauses}, silent intervals within speaker's turns \citep[cf.][]{heldner2010pauses}. With regards \textbf{pauses}, studies of speech corpora show that pause distributions are bimodal or trimodal \citep{demol2006study}, with analyses suggesting that they cluster at around 150 ms, 500 ms, and 1500 ms \citep{campione2002large}, providing empirical support for the traditional classification of pauses into short ($<200  $ ms), medium ($<1000 $ ms) and long (up to $3000 $ ms in spontaneous speech). However, while the nature of the processes that give rise to these classes is poorly understood, what is important for current purposes is that it is possible to make a functional case for this classification.

There is evidence that pauses affect the ease with which speech utterances are processed. Importantly, pauses of different duration affect sentence processing  differently: while pauses below a threshold of around 200 ms are not explicitly detectable \citep{walker1982smooth}, these very short pauses do appear to improve speech recognition. By contrast, the absence of silent intervals or unusually long pauses make sentences harder to comprehend \citep{fors2015production}. More generally, however, while it is clear that they provide crucial information about the relations between events in speech processing, pauses have also been subject of considerable study in relation to learning and error processing in the brain. As highlighted above, feedback adaptation -- an important aspect of articulation -- involves multiple neural systems whose individual contributions are modulated by the durations of intervals between events \citep{teki2011distinct, diedrichsen2005neural, lewis2003distinct, buhusi2005makes, coull2011neuroanatomical}, which in turn affects the quality of the learning outcomes \citep{foerde2011feedback, baese2022just}.

In auditory learning experiments, which like speech perception, involve dynamic reorganization of temporal (when) and auditory (what) expectations through exposure to structured sequences, the specific ranges of interval durations also appear to influence what gets learned. Different delay durations appear to serve as cues to discrimination over the two different sources of information in tone sequences: tone frequency and tone duration. Talking clearly involves more than discriminating between tones of different durations, and speakers seem to be less sensitive to changes in interval duration of speech sequences than tone sequences (and presumably more sensitive to changes in spectral quantities) \citep{grondin2011sensitivity}. However, the mechanisms involved in temporal discrimination of both speech and tone sequences appear to be surprisingly similar, and adaptation to the temporal dynamics of auditory sequences appears critical to speech segmentation and, by consequence, speech intelligibility. Given these parallels, it is notable for current purposes that results from auditory discrimination tasks indicate that inter-event intervals (pauses) shorter than 250 ms and longer than 750 ms only inform temporal discrimination. By contrast, intervals between 250 and 750 ms aid in the discrimination of frequencies as well \citep{buonomano2009influence}. This qualitative divide in the information provided by shorter and longer pauses is also evident in the differences in people's sensitivity to small time perturbations in intervals (both within and between tone sequences). In particular, \textit{just noticeable differences} in durations of intervals shorter than 240 ms are independent of the actual durations themselves, whereas\textit{just noticeable differences} in intervals longer than 240 ms are a function of the durations to be distinguished \citep{friberg1995time, repp2013sensorimotor}.

In relation to speech, the results summarized above indicate that differences in the way pauses from different duration ranges are experienced may affect the way subsequent changes in the signal are detected. Consistent with this, it has been shown that the articulation rate of the earlier parts of utterances co-determines which words and speech segments listeners extract from the speech signal, such that shifting these rates can change what listeners actually hear/extract \citep{dilley2010altering, morrill2014distal, baese2019not}. These effects of speech tempo on speech perception are likely a function of the way variance in the signal influences the way listeners experience the durations of consecutive sounds in sequences. Studies have shown that subjective tone durations stretch or shrink in relation to the durations of the preceding tones in a sequence \citep{nakajima1992time}. Durations of subsequent tones are overestimated if the tones immediately preceding them are considerably longer, whereas tones that follow markedly shorter precedents are experienced as being shorter, with a consequence of this effect being that speakers can experience consecutive sound intervals whose durations are objectively different as being the same \citep{hoopen2006time}. However, the introduction of silent intervals between tones leads to a weakening or an inversion of this effect \citep{sasaki2010time}, a finding that also transfers from the isochronous sequences used in experimental contexts to speech, where pause insertions have been shown to increase the intelligibility of time-compressed speech \citep{ghitza2011linking}.  

Our hypothesis is that alignment is achieved through the establishment of shared expectations about the rate at which informative events will occur, because, by definition, this is a fundamental requirement for speech events to be "informative". Accordingly, we suggest that the intelligibility of speech after linear time-compression \citep{ghitza2011linking} results from a misalignment between the rate at which informative events are expected and the rate at which events actually occur. From this perspective, the reason why pause insertions increase intelligibility is that they serve to counteract erroneous information in the unnaturally compressed preceding part of the signal. This makes the relations between events in the signal more consistent with the temporal expectations speakers will have ordinarily acquired. In other words, the analysis presented above indicates that pauses -- which are most frequently found at the boundaries of the syntactic and prosodic phrases that tend to prompt speech rate transitions \citep{grosjean1979patterns, miller1984articulation} -- may serve to help speakers reset and reorganize their expectations, making changes in the signal informative via systematic alterations to segmentation rate (or phase). 

\section{A theoretical account of the contribution of pauses to speech alignment}

Our hypothesis is that pauses facilitate speaker alignment by making the rate at which informative events happen in the signal predictable and informative. Two simple, unambiguous predictions can be derived from it: first, if pauses serve as an alignment signal, then the distribution(s) of pauses in speech ought to be independent of speaker experience, such that they are identically distributed across speakers. Second, even when speaker expectations are aligned, variations in the rate at which informative events occur will need to be synchronized, and this will be reflected in the way pause distributions vary in time.

\subsection{Formalizing a testable hypothesis}

We next turn to determining the appropriate levels of analysis to apply to these predictions. First, how do we formalize and quantify both convergence and synchronization in the speech signal? We can summarize the requirements of alignment as follows: 
  
\begin{enumerate}
  \item the distribution of pauses ought to be structured so that speakers can extract information from the signal based on a relatively short exposure in time.
  \item the information speakers extract from the signal should be independent of both the individual speakers' experience and fluctuations in pause durations and articulation rates throughout the interviews in the corpus.
\end{enumerate}

In other words, \textbf{local} variation in pause duration when collapsed across speakers in time ought to be statistically independent of the history of the communicative process, regardless of the timeframe used to operationalize history. If the distribution of pauses has this property, then, in theory, at any point in the communicative process, speakers' expectations about this aspect of the signal can be considered to be akin to a 'blank slate'. This critical property for alignment -- that the distribution of some aspect of the time-varying signal be memoryless -- has been shown to apply to discrete communicative events at various levels of description -- i.e. lexical, sublexical, and phrasal -- when they are considered in the communicative contexts in which they are used in text and speech \citep{ramscar2019source, linke2020grammar}. That is, empirically, communicative events approximate geometric distributions at both higher and lower levels of analysis. This applies to linguistic events that are more discrete (e.g. words and phrases), and the more variant sublexical events they provide context for in speech. This finding is important because the geometric is the only discrete distribution that is memoryless. 

Memorylessness describes a formal property of certain distributions of events where differences in knowledge about the events that have occurred prior to a certain point in time confer no advantage. This is because in these distributions, the variance in the rates at which events occur guarantees that knowledge about any events that have already occurred is uninformative in relation to predicting future events. Because the mean of the distribution is always equal to its variance, the variance is bounded by the mean, while the mean is bounded by the distribution of variance. The principle can be formalized in terms of a learning mechanism that seeks to minimize the variable error in the observations with respect to a stable average \citep[cf.][]{peters2022ergodicity}. When a distribution of events is memoryless, it follows that despite any local fluctuations, global event probabilities (or magnitude, in pause duration) are unaffected by knowledge of the history of the process. It thus follows theoretically, that once individuals have experienced a reasonable sample of such a distribution, their models of it will be largely independent of their idiosyncratic experiences. In other words, memoryless distributions ought to provide signals that allow for rapid adaptation of speaker expectations about the rate at which information will arrive.

It follows accordingly that if the distribution of speech pauses is memoryless, then it ought to approximate an exponential distribution (the exponential is the only memoryless distribution for continuous variables). Formally, the exponential is the probability distribution of the time intervals between events in a process in which events occur continuously and independently at a constant average rate. This means that once that speakers have learned to interpret them, local patterns of variation in signals will also have an experience-independent distributional structure. Thus, for present purposes the speech signal can be seen to provide two distinct sources of information: speech pauses which simply vary in duration; and articulated parts of signals that both vary in duration and frequency \textbf{and} which have their own relational structure in sequences. The latter structure reflects the function of speech signals: that they are used to communicate, a process that critically relies on mutual predictability.

\subsubsection{Formalizing context: defining a baseline to measure misalignment in time}

The degree to which signals are predictable will be influenced by the amount of information available prior to a given point: their context. This raises a problem when we consider the communicative constraints speech codes must satisfy, namely that the experiences of the users of a code will vary greatly, both in what they have sampled and how often. This suggests that across speakers, individual ability to segment the signal and then use this to model context will vary. Moreover, these apparently separate processes will interact, such that as a result of this interaction both what a 'segment' is -- and in particular, the amount of information required for its presence to be inferred -- and the amount of information that 'segment' in turn contributes to contexts will both change with experience.

It follows that because articulations vary both with context and individuals' experience over time, they cannot themselves provide a baseline for our theoretical analysis of pause distributions. Rather, to test our hypothesis we require a signal dimension that develops linearly over time. One source of such information is sequence position. Sequence -- or utterance -- length is determined by variation in structural regularities at various levels, including word order, word morphology, and argument patterns. The rates at which regularities at these first two levels -- which roughly correspond to syllable and short word boundaries -- are produced appear to be relatively invariant both in terms of development of individual speakers and different languages \citep{poeppel2020speech, luo2007phase}. By contrast, sequence-level patterns of development at any timescale are characterized by distributional changes that involve an increase both in the length of sequences and the frequency of shorter sequences. Accordingly, we will use sequence position and its pattern of development with experience as a baseline in our analysis. We will examine the relationship between this baseline and our hypothesized time invariant source of information provided by pauses and analyze the predicted changes this leads to with increasing speaker experience. 

\subsubsection{The hypothesis: predictable divergence between distinct signal dimensions over time}

If pauses provide information about the rate at which other informative events occur in signals, and if experience changes the rate at which events occur (and hence even their nature), it follows that the relationship between pauses and shared rates (i.e., word and syllable boundaries) ought to change at those points that are most affected by experience. 

Lifelong experience is typically accompanied by some form of specialization. Adulthood typically involves the pursuit of different vocational paths and different interests. These inevitably increase both context-specific knowledge and vocabulary \citep{ramscar2014myth}, which ought to be reflected in the production of signals that are increasingly context-specific, that we expect in turn to interact with sequence position. This prediction can be explained as follows: we have described above how communication of context-specific knowledge relies on -- and hence will develop through -- a transmission process that requires the maintenance of mutual predictability. We have suggested that this is achieved by the systematic adaptation of phrase structure and sequence length, a process that, at the utterance level, serves to reduce communicative uncertainty across structured sequences. Accordingly, it follows that if segmentation rate increases are a function of the relative predictability of the structured sequence, experience ought to lead to an increase in segmentation rate and a decrease in segmentation rate variability in later sequence positions.

A further implication of the interaction between experience, sequence position, and segmentation rate is that because context-specific signals are irregularly distributed themselves, experience will also lead to increases in the variability of local segmentation rates -- bursts of activity -- as opposed to global increases that would spread out uniformly across older speakers' signals. Notably, the correlation between information rates and signal sparsity appears to be an instance of a more general phenomenon. Bursty patterns of activity are a characteristic of complex event dynamics, such as for example traffic and internet activity \citep{karsai2018bursty}, and the way words occur in language use \citep{katz1996distribution, altmann2009beyond}. Which brings us back to the analogy between intervals between general classes of events and speech pauses. While gaps between events are a well-known quantity in the analysis of complex event dynamics, to our knowledge intervals between events have not previously been considered a source of information in relation to distributions of automatized complex cognitive behaviors. Yet, given that it has been shown that fluctuations in inter-event times are a better predictor of irregular bursts of activity characteristic to complex behaviors in humans than the event distributions themselves \citep{goh2008burstiness}, there is reason to suppose that they might also provide information about the processes that give rise to these behaviors.

As reviewed in section 2, pauses from different ranges of duration modulate the extent to which exposure to acoustic signals leads to temporal adaptation, and whether this exposure also results in adaptation to sensory prediction error. These two signal dimensions - timing and the information provided by the frequency spectrum - contribute to distinct aspects of behavior: first, uncertainty management, which involves coordination of expectations in time; second, permanent adaptation in patterns of execution (e.g., the way speech signals are produced in context). Both of these mechanisms involve learning and take time to develop. The extent to which speakers' ability to utilize them efficiently changes with experience is determined by the structure of the speech samples they are exposed to, and the order in which distinct aspects of the signal become predictable and thus informative. At phrase level, for example, it is less likely that adult speakers will encounter novel verb argument structures. By contrast, the number of nouns they encounter within these structures will increase steadily across the lifespan \citep{ramscar2014myth}. As a consequence, the information provided by verb argument structures will decrease in relation to the increasing number of context-specific nouns conditioned on them. More generally, experience will lead to an increasing asymmetry between variability (and uncertainty) in some aspects of the signal (e.g., context-specific articulations) in relation to the relative invariability of the communicative contexts they are conditioned on \citep{ramscar2021children}. Simultaneously, more experienced speakers will acquire increasing certainty about those aspects of the signal/environment that are relatively invariant in time -- e.g., how to structure sequences for successful uncertainty management and efficient message transmission.

These theoretic assumptions allow us to formalize two key questions we shall now address: First, are speech pauses distributed so as to allow speaker alignment in the way we suggest? Second, are the interactions between pause duration, timing, and speaker experience consistent with our predictions? If speech sequences are structured for incremental uncertainty reduction, the average information rate should increase with the relative sequence position (order). This means that information in speech ought not to be uniformly distributed in sequences (as indeed is the case \citep{genzel2002entropy, genzel2003variation}).  We suggested above that this asymmetry in the temporal distribution of information in speech sequences will increase with speaker experience, increasing the differences between the rates at which individual speakers segment the signal. We hypothesize that speech pauses serve to offset these differences and help speakers organize their expectations about the rate at which information arrives in time. These considerations yield three concrete predictions that we will test in the analysis that follows: 

\begin{enumerate}
  \item That the aggregated samples of pause durations will converge on the exponential distribution.
  \item That experience will optimize speech production by local adaptation of signal variability, which will lead to an increase in information density ('segmentation rate') in the utterance final positions of longer utterances. That, in final positions of longer utterances, the uncertainty reduction by the previous part of the utterance allows information to be transmitted at higher rates, and a decrease in segmentation rate in the utterance initial positions and short utterances, where the uncertainty is high and information rates are low. This in turn will lead to an increase in signal sparsity and increasingly bursty patterns of activity in pause production.
  \item That this tendency ought to be reflected in a systematic divergence from the mean pause duration with phrase length and the proximity of the phrase boundaries in speech produced by older speakers -- i.e., we should expect more extreme values at the boundaries of both very short and very long phrases.
\end{enumerate}

\subsection{Corpus Data}

We analyzed pauses from the Buckeye Corpus of Conversational English \citep{pitt2005buckeye} and the Korean Corpus of Spontaneous Speech \citep{yun2015korean}. The corpora each contain phonetically transcribed speech from informal interviews with 40 speakers balanced by age and gender. The young cohort in the English corpus consists of speakers below the age of 30, the older speakers are aged 40 and upwards, with more age variance in the older speaker group. The Korean speakers are aged 15 to 47 years, with a median of 29. 

All phrases produced by the interviewees were manually transcribed and word and phone boundaries were annotated with an automated speech aligner. Subsequently, the aligner annotations were manually corrected by trained annotators. Periods of silence that occurred between word boundaries were tagged as pauses. Absences of articulation accompanied by audible breathing or other sound were tagged separately. Short periods of silence that occurred between word boundaries were labeled as silent phones, stop closures, or assigned to surrounding phones \citep{kiesling2006variation}. For the purposes of the analyses reported here, we focus exclusively on silent pauses between words.

The Buckeye corpus contains $26952$ and the Korean corpus $21571$ pauses. 98.7\% of the pauses in the English sample and 99.75 \% in the Korean sample are shorter than 3000 ms. To facilitate the analysis of pause duration in relation to utterance position, each pause was assigned to the utterance position of the word it preceded.

We examined the interaction of pause distribution, segmentation rate variability, and experience, using two parameters: utterance position and age cohort. Compared to the Korean sample, the English sample provides a larger span of speaker ages and concomitantly speaker experience. This enables us to analyze speech data that allows for a strong test of this hypothesis. Previous findings have shown that response differences between age cohorts can be explained by the systematic changes of co-variation patterns in the samples speakers are exposed to and learn from across the lifespan and that the onset of these 'aging effects' is measurable in speakers' performance relatively early (in their twenties, see Figure~\ref{fig:age}, panel D) \citep{ramscar2013learning}. Therefore, despite the differences in the age range, we can reasonably expect to find measurable differences in the performance of the 'older' group of speakers (speakers above the age of 30) in the Korean sample.

\section{Results}

For the pauses extracted from the English and the Korean corpus we estimate the rate parameter $\lambda$, and model the exponential distribution using maximum likelihood estimation (for details, see Supplementary Materials).

Figure~\ref{fig:distributions} shows the model and empirical distribution with best fits to the target range of durations (0-3000 ms). The empirical distribution of speech pauses in both English and Korean approximates the exponential, showing some divergence in the high-frequency range that hosts the shortest pauses ($<100 ms $) and in the tail of the distribution, where the random sample model fit predicts a larger range of values (extremely rare pauses longer than 3000 ms). The model distributions fitted to the selected range of values, such that the simulation approximates the best fit within the range (instead of the best fit to the number of events imposed by the corpus size), and the empirical distribution shows a close fit to the exponential model and the truncated random sample of exponentially distributed values. 

To test for the distance between the model and empirical distribution we performed a two-sample Kolmogorov-Smirnov test. The test shows a relatively small, but significant difference between the empirical distribution and the truncated model, in both languages ($D_{ks}=0.09, p<0$). It seems worth noting here that the test is very sensitive to small differences between distributions and also shows a significant difference between the truncated exponential and the exponential model ($D_{ks}=0.01, p<0$). The distance between the empirical distribution and the model reflects the misalignment in the high-frequency part of the distribution that holds the shortest pauses. This is  unsurprising given that unvoiced stops and silent vowels (whose durations vary in the range of 20-100 ms) were excluded from the analysis. (Detailed test statistics and discussion of alternative distributional hypotheses are provided in the supplementary materials.)

\begin{figure}[ht]

\centering
  \includegraphics[width=1\textwidth]{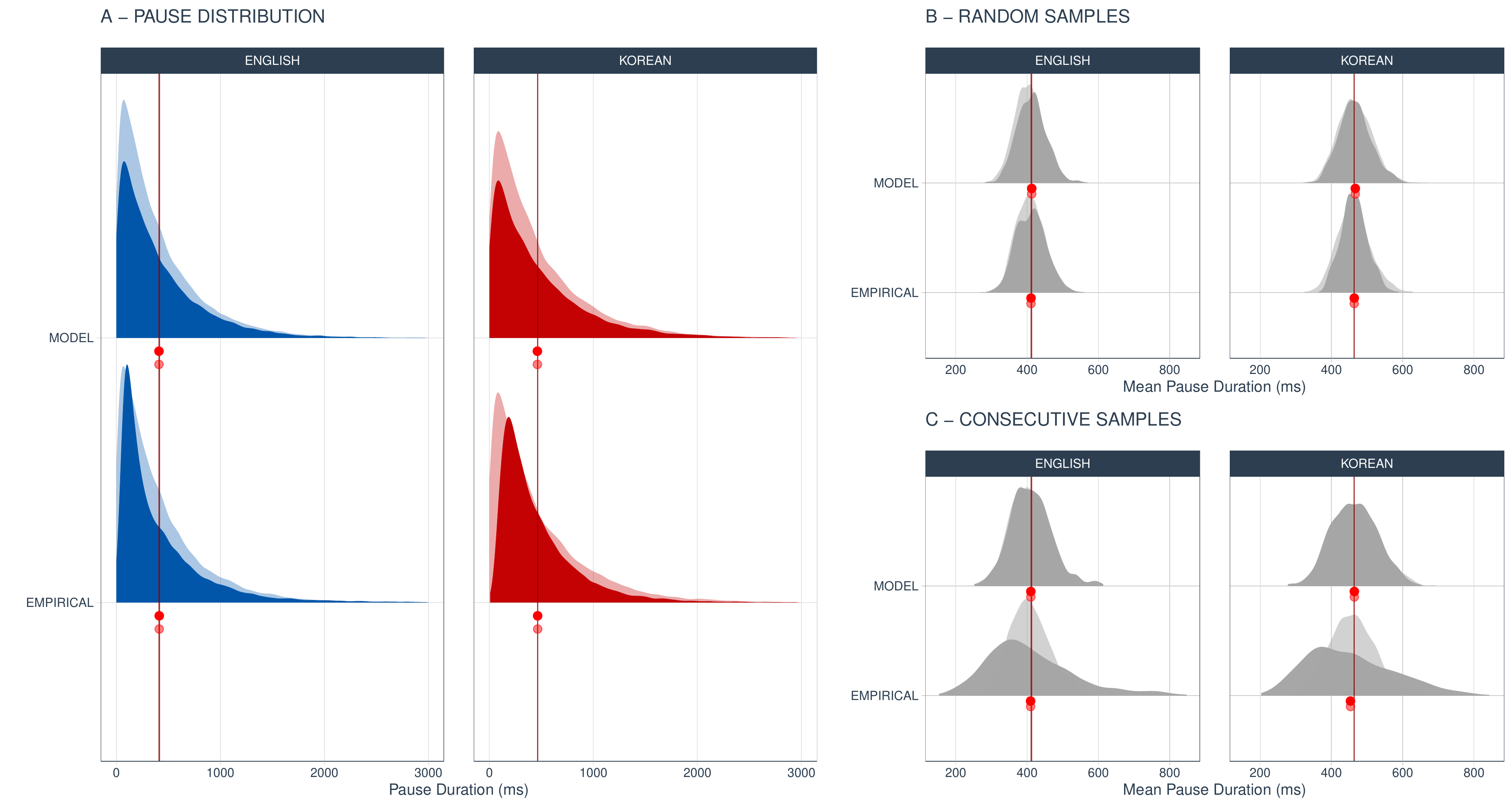}
 \caption{Left panel (\textbf{A}): probability density function of pause distribution for pauses shorter than 3000 ms (bottom row) and a random sample from an exponential distribution (top row) of identical size and rate parameter. The shaded area shows the truncated exponential model, limited in the analyzed range of values $(0,3000)$. The sample means (red dot) and model mean (transparent red dot) center at the theoretical mean (red line). Right panel: the distribution of sample means in samples ($n = 50 $) drawn from the distributions shown on the left. Means from random samples (\textbf{B}) converge on a normal distribution that centers at the theoretical mean. Means from samples of consecutive events (\textbf{C}) show the same behavior for model distributions, while means from consecutive samples drawn from the empirical data are both more dispersed and left-skewed, indicating local biases in the distribution of pause durations.}
 \label{fig:distributions}

\end{figure}

Consistent with previous findings \citep{linke2020grammar}, individual speakers' samples seem to deviate randomly, with samples approximating the gamma distribution with varying scale parameters. The aggregate approaches the exponential distribution (which is a special case of gamma). This means that despite a multitude of sources that concomitantly feed into individual variation (i.e. individual experience, syntactic structure, and various other factors that shape interactions and rates at which acoustic events are extracted from signals) the aggregate distribution maintains a structure that can allow speakers to converge on a stable mean expectation. To examine the effects this individual variation has on the local distribution of variance, we run a sampling simulation. In order to investigate whether speech pauses exhibit bursty patterns of activity, we draw 1000 random samples of 50 'pauses' from each distribution and samples of 50 consecutive 'pauses', comparing the distribution of mean duration in relation to the theoretical mean $1/\lambda $. The right panel shows the distribution of averages from random samples (top) and sequences (bottom) for Korean (right column) and English (left column). Sample averages from random samples and samples of consecutive events from the model distribution converge on a Gaussian distribution that centers at the theoretical mean. We observe more variance in the means of sequential samples. Samples of consecutive events drawn from the empirical distribution, show a strong left skew and more dispersion, suggesting that extreme values in pause durations are not distributed uniformly in speech sequences.

These results suggest that pause distributions possess two desirable properties: one, a time-invariant global distribution that can provide developing speakers with a consistent source of information in the otherwise noisy signal they are exposed to; two, systematic local fluctuations in average pause duration that can allow adult speakers at different levels of experience to rapidly adapt to changes in local segmentation rates given relatively short signal samples. All of which raises a critical question: are local fluctuations in mean pause durations systematic in relation to changes in the distribution of information (and uncertainty) in signals? 

\begin{table}

\centering
{\begin{tabular}{lcccccc} \toprule
 & \multicolumn{2}{l}{RANDOM} \\ \cmidrule{2-7}
 ENGLISH & Mean & Median & SD & SE & IOD\footnote{1} & CV\footnote{2}(\%) \\ \midrule
 model & 407.63 & 407.35 & 60.26 & 1.91 & 8.91 & 14.78 \\ 
 empirical & 409.96 & 405.86 & 59.85 & 1.89 & 8.74 & 14.60 \\ \\ 
 KOREAN & & & & & &  \\ \midrule
 model & 468.13 & 463.06 & 67.96 & 2.15 & 9.87 & 14.51 \\ 
 empirical & 464.14 & 461.25 & 51.89 & 1.64 & \textbf{5.80} & 11.18\\ \\ \midrule
 & \multicolumn{2}{l}{CONSECUTIVE} \\ \cmidrule{2-7}
 ENGLISH & Mean & Median & SD & SE & IOD & CV(\%) \\ \midrule
 model & 412.40 & 409.51 & 54.77 & 1.73 & 7.27 & 13.28 \\ 
 empirical & 412.12 & 390.90 & 126.83 & 4.01 & \textbf{39.03} & 30.78 \\ \\ 
 KOREAN & & & & & &  \\ \midrule
 model & 466.75 & 464.31 & 61.56 & 1.95 & 8.12 & 13.19 \\ 
 empirical & 468.23 & 447.11 & 130.32 & 4.12 & \textbf{36.27} & 27.83 \\ \\ \bottomrule
 \multicolumn{7}{l}{\textsuperscript{1}Index of Dispersion, \textsuperscript{2}Coefficient of Variance (\%)} \\
\end{tabular}}

\caption{\textbf{Distribution of pause duration from samples of conversational English and Korean.} Mean durations in random samples drawn from the empirical distribution of Korean pauses are less dispersed than the model mean distribution. As shown in Figure 1 of the results section of our article, samples of consecutive pauses from the empirical distribution are more dispersed than consecutive samples from the model distribution. This indicates that pauses from different ranges of duration are not uniformly distributed in time, and instead appear in local bursts.}

\label{sample-table}
\end{table}

\subsubsection*{How does pause duration interact with the event rate? What is the event rate?}

Formally, the exponential is a distribution of waiting times between events generated by a Poisson process. The $\lambda$ parameter defines the rate at which these 'events' occur: the average number of 'events' within a fixed time interval. We discussed earlier how in language, the informative changes in signal dimensions that can constitute informative 'event' boundaries shift more or less systematically with speaker experience. As a consequence, the rates at which the respective 'events' occur ought to shift more or less systematically too. It follows therefore that rate estimates only ever approximate the convergence rate, and do not represent fixed, objective estimates of the rate at which individuals will extract informative changes from the signal. 

With respect to the pause distributions, which as we have shown in both Korean and English approximate the exponential distribution, the rate parameters ($\lambda_{ko}=0.0022$, $\lambda_{en}=0.0024$) indicate that within a fixed frame of $1000$ ms, Korean speakers ought to converge on a 'minimal agreement' that informative changes in the signal can be 'expected' to occur every $464$ ms ($1/\lambda$)  on average and that, given that pause likelihood decreases exponentially with pause duration, the probability of subsequent acoustic events occurring increases exponentially with pause duration. In other words, if pauses are information and information processing is probabilistic, an exponential decrease in probability as a function of pause duration indicates that there are large differences in the way small perturbations in pauses from different ranges of duration interact with probability. 

In this model, the probability of an informative change (articulation) increases considerably as pause duration increases from 100 ms to 150 ms, while hearing more silence is almost equally unexpected between 750 to 800 ms. This in turn provides an interesting probabilistic perspective on the possible role of pause duration in the dynamics of uncertainty management, and the response to informative changes in articulations. It suggests that all articulations following longer pauses are uninformative because they are over-expected, while the informativeness of articulations following short pauses will vary with the expectations built up by the previous part of the articulated signal. If we apply the same reasoning to pauses from the middle range, then the information provided by subsequent articulations ought to be a compromise function of both pause probability and the articulated prior. With respect to pauses in speech, and the way they manage expectations (i.e., the information rate), an over-expected stretch of speech signal immediately following a long pause can be conceptualized as follows:

\begin{itemize}
     \item[] [1269 ms PAUSE] \textit{that} a president would do something ...
\end{itemize}

\noindent
In this example, the pause duration leads to an expectation of the way \textit{that} will be signaled. In order to be intelligible in isolation, \textit{that} would ordinarily be signaled using multiple, discriminable acoustic segments \textit{dh ae t}. However, in the context of the long pause that precedes it, the measurable acoustic-phonetic information it contains can deviate from the isolated signal as long as the information it contributes does not violate speakers' expectations about what \textbf{can} follow.  That is, at points where articulated signals are over-expected, the only thing that will violate expectations is the absence of articulation -- hearing any articulation is more probable than hearing more silence. Simultaneously, speakers' expectations about the way a signal following a pause will unfold in time (where word and syllable boundaries occur) appear to be relatively fixed. This means that whatever is articulated instead of \textit{dh ae t} -- even if it is a mere burst of noise -- must occur within the expected time interval in order to maintain the intelligibility of those parts of the signal that follow it  \citep[see][]{doelling2014acoustic}. 

At the other end of the pause spectrum, the fine-grained acoustic detail that follows the bulk of short pauses ought to become increasingly informative and diversified in signals in which they occur. Because short pauses provide little useful information on their own, listeners ought to get better at noticing whether any violation of prior expectation actually occurred in the acoustic signal and respond with fine-grain temporal adaptation. From this perspective, the articulated signals that are learned and remembered are systematic independent of speaker experience, because they are shaped by the pause distribution and vary consistently with pause duration.

Accordingly, it follows that if the exponential model applies to speech pause distributions in the way we suggest, then segmentation rates ought to become increasingly diversified with experience. To investigate this question we analyze experience-related change in utterance structure and the distribution of pauses.

\subsection{The interaction between sequence length, pause duration, and experience}

We analyze the frequency distribution of utterance positions and pause durations. Prior analyses have shown that the distribution of words across utterance position, and word length in phonemes in speech data follows the geometric distribution. This indicates that segmentation rates that approximate word and phoneme boundaries in spontaneously produced signals vary consistently in time. The correlation between log-transformed frequency and frequency rank in these distributions reflects their fit to the geometric distribution.

\begin{figure}[ht]

\centering
 \includegraphics[width=0.99\textwidth]{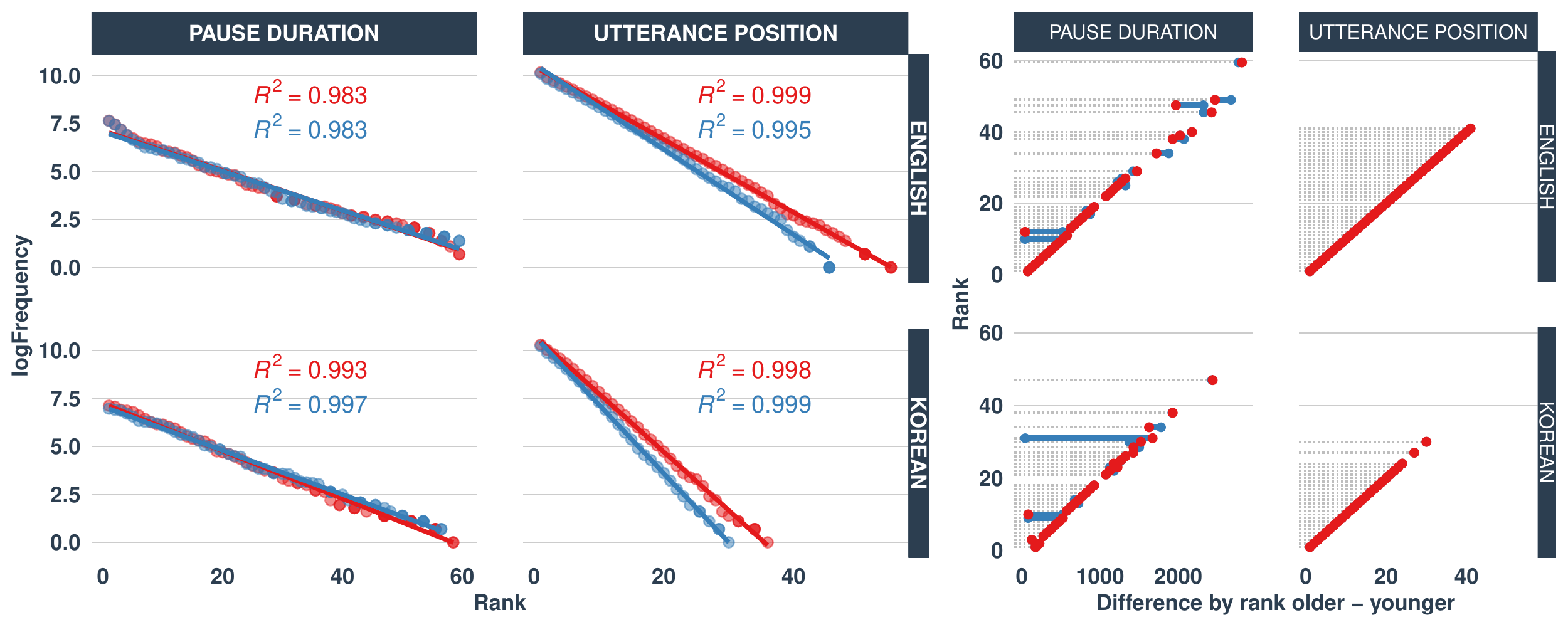}
    \caption{{Left panel: Frequency distributions of utterance position (left) and binned speech pause duration (right, bin size 50 ms) for younger (blue) and older (red) speakers. Lines show model fits to geometric (linear after half log transformation). Both the pause distribution and the utterance position distribution show close fits to geometric with and $R^{2}>0.99 $respectively. Note that the lines that represent the geometric model distribution are hard to distinguish because the distribution and the model fits are nearly identical. In the left column, the younger and the older cohorts' distribution and their respective fits overlap completely. Right panel: point-wise correlations of pause duration (left) ranked by frequency show some minimal misalignment in the rank distribution. As can be seen in the rightmost column, the ranked distributions of utterance positions are identical. Older speakers produce longer utterances on average while maintaining the shape of the distribution and the relationship between utterance length and utterance probability. This correlation is unusual, when same analyses are performed on text, the correlation between ranked values from any two samples in a mixed corpus tend to be weaker.}}
 \label{fig:information}
 
\end{figure}

The analyses presented in Figure~\ref{fig:information} show that frequency of utterance positions closely approximate a geometric distribution in both languages and cohorts ($R^{2}_{en}>0.994 $, $R^{2}_{ko}>0.997 $). While Korean utterances are on average shorter ($M=4.017$, $SD=3.302$, $n=57661$) than English utterances ($M=5.710$, $SD=5.357$, $n=50841$), older speakers of both languages produce longer utterances more often (Mann-Whitney, Korean: $M_{o}=4.37, M_{y}=3.74, U=0.47, p<.0032$, English: $M_{o}=6.13, M_{y}=5.50, U=0.28, p<.0763$). The medians are significantly different in older speakers of English (Mood, Korean: $Mdn_{o}=4.14, Mdn_{y}=3.64, Z=-1.13, p-.2573$, English: $Mdn_{o}=6.20, Mdn_{y}=5.65, Z=2.43, p<.015$), indicating divergence in scale and more individual variability in utterance length in older speakers of English. Note that we are not interested in individual variation in the analyses reported here, rather the analysis explicitly targets aggregate behavior. The results show that the older speakers maintain the optimal distribution \textbf{despite} the increase in variability across individuals. (All descriptive statistics and results of statistical tests are provided in the supplementary materials.)


As discussed above, spontaneous behavior can be expected to vary across individuals, and with experience, the variance in the articulated parts of signals can be expected to increase. Accordingly, we examined whether the variance in individual behavior does in fact systematically increase in the aggregate (despite the many sources of variation that shape individual behavior in seemingly unsystematic ways). As expected, the distributions maintain their shape (they stay geometric) and as a result, the slope of the older speakers' distribution decreases. The changes in slope indicate a decrease in the information provided by the relationships between words, i.e., their relative probabilities in context. Typically, decreases in performance variability over a finite set of outcomes like this are taken as a mark of learning, and the  decrease in noise to signal ratio associated with it \citep[cf.][]{shmuelof2014recent, tucha2004handwriting, herman2009saccade}.

This course of development is consistent with the idea that changes in the signals produced by older speakers reflect the patterns of learning that can be expected to accompany increased experience rather than any 'pathological' decline in performance \citep[cf.][]{ramscar2014myth, campbell2016robust}. The decrease in structural uncertainty (i.e., the amount of variance in the conditional probabilities among adjacent words) in older speakers facilitates a redistribution of functional load: the uncertainty at word transitions decreases, freeing up resources to allow learning of fine-grained, context-specific patterns of articulation \citep[cf.][]{poulisse2020oscillatory}.

\begin{figure}[ht]

\centering
 \includegraphics[width=0.99\textwidth]{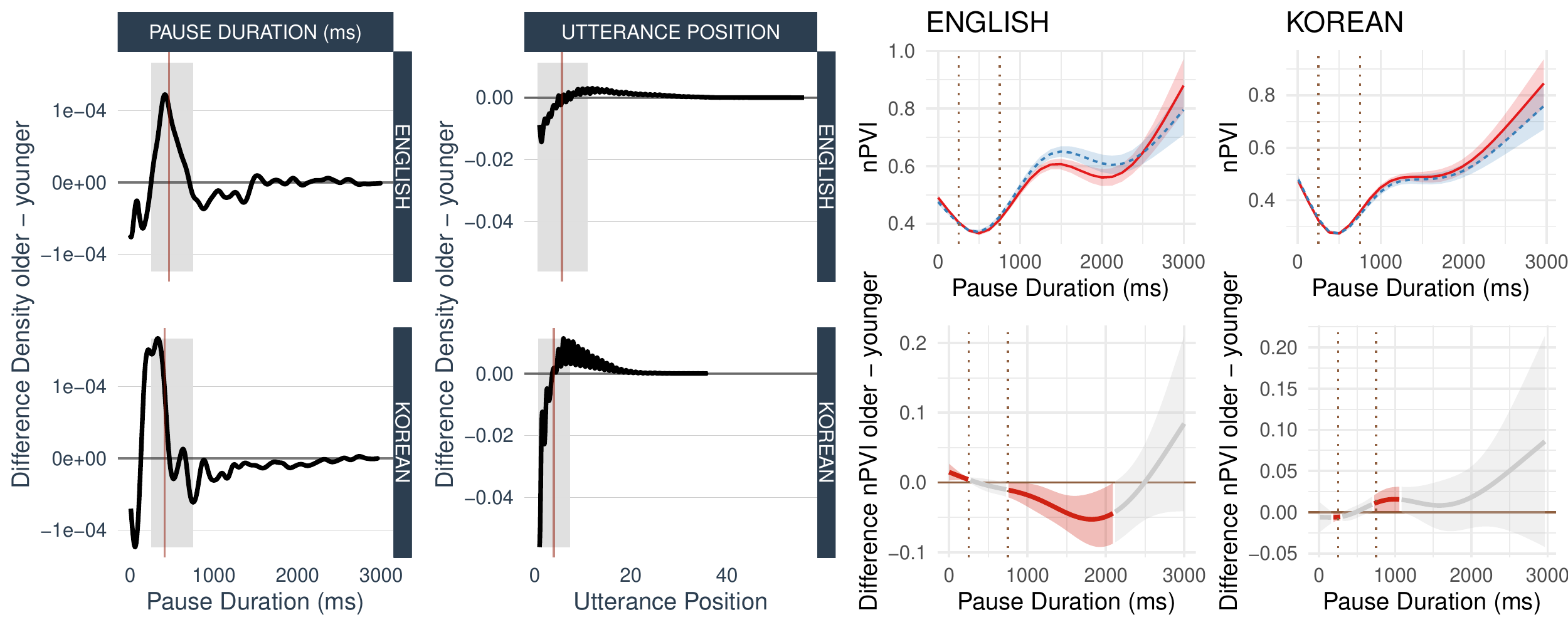}
 \caption{\textbf{Left panel:} Difference in probability density of utterance position (right panel) and pause duration (left panel) in speech samples produced by older speakers of English (top row) and Korean (bottom row). The red line marks the sample mean, the gray area highlights pauses from the middle range of durations (250-750 ms) and the standard deviation from the mean in the probability of utterance position. As can be seen, plots reveal a shift towards the mean (convergence) in pauses shorter than 1000 ms, indicating that longer than average pauses become shorter and shorter than average pauses become longer in older speakers, decreasing the individual variance in pause duration. By contrast, for utterance position experience shifts the average towards the right showing an increase in variability (i.e, while pauses converge, utterances appear to diverge). The effect is more protracted across utterance positions in English speakers, which is consistent with the longer average utterance length in English. Right panel: Pairwise pause variability (\textbf{nPVI}) as a function of pause duration and speaker age. The top panel shows the relationship between pause duration and pairwise variability, which is identically u-shaped in both languages, reaching its minimum at the mean pause duration. The bottom right panel shows the differences between cohorts (areas, where the difference is significantly larger than 0, are highlighted in red), which show opposite patterns in Korean and English.}
 \label{fig:density}

\end{figure}

\begin{figure}[ht]

\centering
 \includegraphics[width=0.99\textwidth]{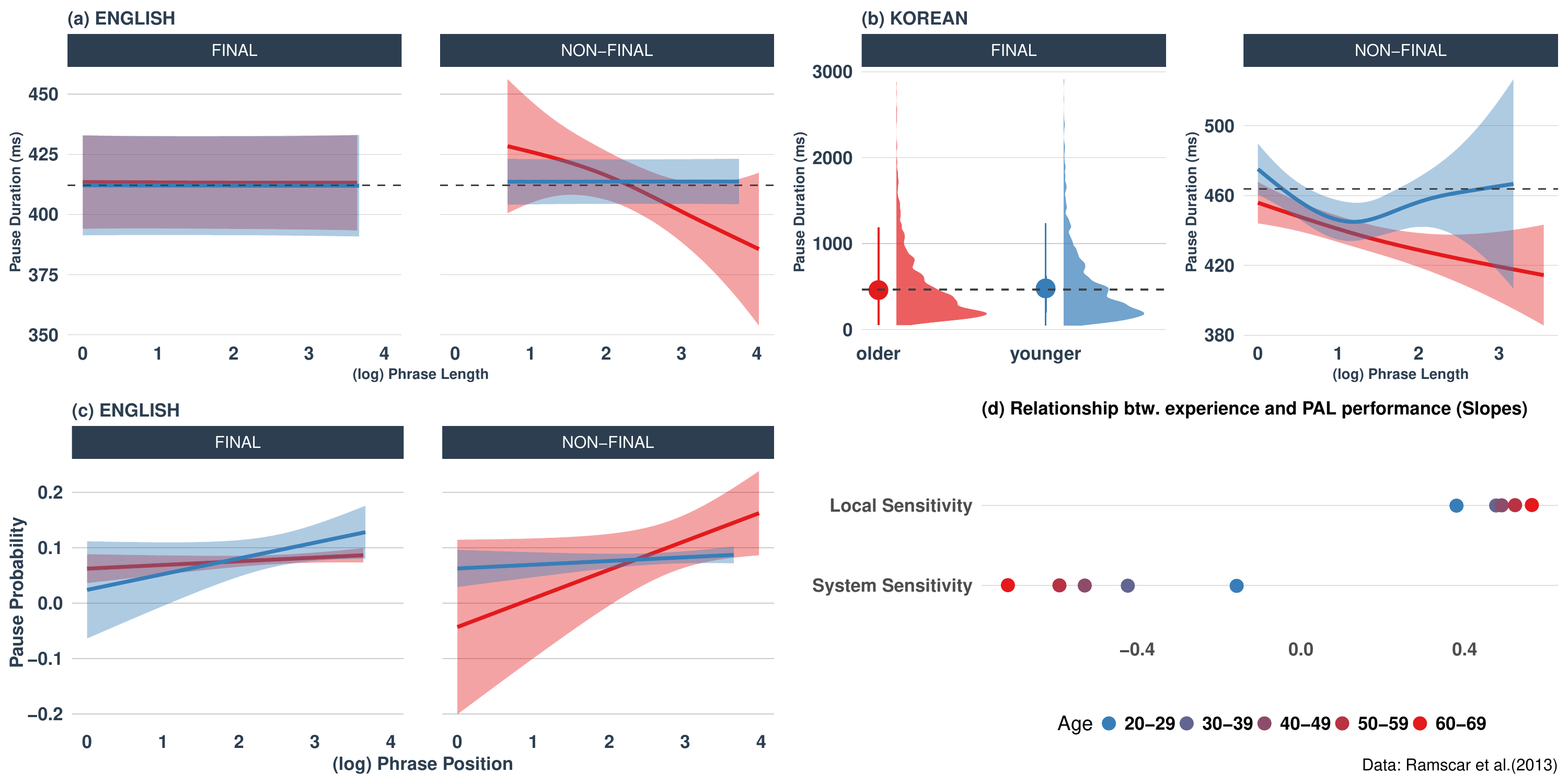}
  \caption{Mean pause duration as a function of utterance length for older(red) and younger (blue) speakers of English (a, top left) and Korean (b, top right). In the Korean sample, pauses occur at phrase initial boundaries only, pauses in the English sample are distributed across the utterance - the distribution of utterance position of words preceded by pauses is geometric. Pause probability decreases with utterance position in both younger and older speakers, but pause probability appears to shift towards the later positions in older speakers (panel c): In English, experience increases the average duration of pauses preceding non-final words in shorter sequences and decreases the average pause duration in longer sequences. In Korean, the cohorts diverge in pause durations preceding longer utterances only but generally appear to model a similar behavior, which becomes less noisy with experience -- older speakers produce shorter pauses, and there is less variance in duration. \textbf{Panel d} shows the model expectation: effect of experience-related changes in English speakers' sensitivity to the frequency of co-occurrence relationship (system sensitivity) and cue frequency (local sensitivity) on performance on the paired associate learning task by age group.}
 \label{fig:age}

\end{figure}

However, whereas we observe a marked effect of experience on the articulated part of signals, it does not appear to affect pause distributions. The frequency distributions of pauses (with pause duration binned at 10, 20, or 50 ms) in both age groups closely approximate the geometric ($R^{2}_{en}>0.984 $, $R^{2}_{ko}>0.993 $) and both distributions have identical slopes. Simultaneously, the mean pause duration decreases in the older cohorts. The differences in the means are not significant (Mann-Whitney, Korean: $M_{o}=439.28, M_{y}=454.96, U=0.1, p=.5291$, English: $M_{o}=400.88, M_{y}=408.21, U=0.0698, p=.6588$), and neither are the differences in the medians (Mood, Korean: $Mdn_{o}=419, Mdn_{y}=441, Z=-0.95, p=.3398$, English: $Mdn_{o}=381, Mdn_{y}=413, Z=-0.12, p=.9081$). In English, we observe more variability across older speakers and a decrease in variance in the aggregate. (See Supplementary materials for a full summary of descriptive statistics.) A closer inspection of differences in the probability density (Figure~\ref{fig:density}) reveals that older speakers of both languages produce more pauses from the middle range of durations (250-750 ms). In other words, older speakers of both languages produce fewer pauses that markedly diverge from the mean duration. This suggests that in pause production, experience leads to a gradual convergence on the mean.

To summarize these findings, experience appears to result in a redistribution of uncertainty across utterance positions as it increases. The differences between the distributions indicate that experience-related changes in the distribution of uncertainty in the articulated parts of signals increase linearly with utterance position. The fact that the distribution of pause durations does not change with experience indicates that the relationship between pause duration, utterance length, and uncertainty will change systematically over time. The slopes in the distribution of utterance positions decrease, indicating a decrease in differences in the distribution of uncertainty across utterances. This implies that if pause durations reflect uncertainty, the relative differences between consecutive pauses should also decrease with utterance length. This also suggests that alignment in communication relies on two distinct notions of convergence in probability. Both alignment and stochastic convergence rest on the idea that a stream of random or unpredictable events or quantities (noise) can settle into a predictable behavior over time. A predictable behavior can be characterized by: 

\begin{enumerate}
  \item a decrease in contrasts between consecutive values (the observed events or quantities eventually become indistinguishable)
  \item a stable probability distribution, where the contrast between consecutive values is maintained, but the change is kept predictable (which can be achieved by learning to ignore unpredictable variation)
\end{enumerate}

We propose that communication between speakers at different levels of experience relies on both of these different notions of convergence and that the degree to which speakers utilize one or the other changes across the lifespan. So far, we have described these distinct aspects of alignment as structural or systemic (pertaining to relationships between informative events) and contrastive (pertaining to discriminable changes in the informative events themselves). Consistent with this theoretical analysis, these results suggest that experience acts as a 'sink' in that it decreases the variability of higher-level structural or relational information, eventually making highly predictable events indistinguishable.

To further examine this, we next analyzed the effect of experience on the relationships between consecutive pauses and their interaction with pause duration. Changes in the relationships between consecutive pauses indicate uncertainty changes and the rate at which the signal is segmented. Larger differences ought to indicate larger changes in segmentation rates of consecutive articulation. That is the deviation from the 'event' or segmentation rate in the articulated stretch of signal that separates two pauses ought to be reflected in the durational contrast, such that convergence in rate, or an increasingly uniform distribution of information, mediated by the convergence in sequential order, would result in a decrease in durational contrast between consecutive pauses.

\subsection{Changes in durational contrast between consecutive pauses}

To quantify these changes in durational contrasts, we calculate the normalized pairwise variability index (\textbf{nPVI}) of adjacent pauses (pairs of pauses separated by an articulation). This metric quantifies the average durational contrast between consecutive events in acoustic sequences (e.g., notes in a melody or vowels in speech utterances). It was originally developed to distinguish timing patterns in languages that differ in prosodic structure \citep{ling2000q, grabe2008durational} and captures the relative variability of segment durations in relation to the event onsets of isochronous sequences. The nPVI for a pause occurring at position \textit{k} and \textit{k+1} in a sample of \textit{m} silent pauses of duration \textit{d} is calculated as $$ \sum_{k=1}^{m-1} \frac{|d_k - d_{k+1}|}{(d_k + d_{k+1})/2} / (m-1)$$ The metric allows us to estimate differences in the duration of adjacent pauses independent of local fluctuations in the articulation rate. The nPVI ought to reveal changes in the temporal relationships independent of the pause durations themselves. i.e., if pauses do inform timing, relative changes in duration provide information about changes in the segmentation rates of articulated signals that separate two consecutive pauses, independent of context-related variability of segmentation rates. 

We observe a decrease in the average nPVI in older speakers of both languages (Mann-Whitney, Korean: $M_{o}=.336, M_{y}=.345, U=.154, p=.327$, English: $M_{o}=.425, M_{y}=.430, U=1.462, p=.713$), which could indicate a decrease in rate variability in the articulated parts of the signal that constitute the transitions between consecutive pauses. Moreover, there is less variability in temporal contrast between adjacent pauses in Korean (Mann-Whitney: $M_{ko}=.341, M_{en}=.427, U=1.462, p<0$), a trend similar to those previously observed in vowel durations \citep{grabe2008durational}.

We model the relationship between pause duration and the nPVI as a non-linear two-way interaction with experience (Figure~\ref{fig:density}, left panel) in a generalized additive model using the R library \textit{mgcv}. Generalized additive models allow us to examine non-linear interactions between multiple predictor variables while accounting for the highly skewed distribution. We fit a Gamma model with a log link function. The model is specified as follows: $$nPVI \sim s(PauseDuration, by=cohort, k = 5)$$ We find a u-shaped effect of pause duration on durational contrast in English (older: $edf=3.977$, $F=187.6$, $p<0$, younger: $edf=3.976$, $F=245.9$, $p<0$) and Korean (older: $edf=3.978$, $F=207.8$, $p<0$, younger: $edf=3.976$, $F=176.0$, $p<0$). To test for differences between the cohorts, we examine the difference curve between the smooths of the two factor levels (older and younger speakers). The relationship between pause duration and durational contrast to adjacent pauses does not differ significantly between the cohorts in pauses from the middle range of durations, whereas cohort differences in pauses shorter than 250 ms and longer than 750 ms are significant (Figure~\ref{fig:density}, the bottom left panel shows the difference between the smooths of the two cohorts, the factor levels 'older' and 'younger', parts of confidence intervals that do not include 0 are highlighted in red). In pauses produced by older Korean speakers, there is an increase in durational contrast between shorter pauses and their neighbors, and a decrease in longer pauses. By contrast, older speakers of English decrease the durational contrast between longer pauses and successors and increase it in shorter pauses. 

To test whether the differences in the duration of pauses produced by older speakers is correlated to the increasing sparsity in the distribution of information in the way we suggest, we next model the probability and duration of pauses produced by younger and older speakers as a function of utterance length and utterance position.

\subsection{Does experience change the relationship between pause and utterance length?}

In the English sample, pauses are distributed across utterances and the distributions of utterance positions preceded by pauses are geometric; they do not differ from the overall distribution in both cohorts. In Korean, silent pauses are found at the utterance initial boundary only. This is consistent with the differences in the prosodic structure of Korean, which tends towards isochronous syllable timing \citep{tark2012experimental, moon2004rhythm}, and English, which is stress-timed and highly variable in syllable timing.

We fit a generalized additive model of pause duration as a smooth function of utterance length for each cohort, adding a factor term for the utterance final boundary (a categorial predictor to distinguish between final and non-final positions of utterances). We use a Gamma model with a log link function to account for the highly skewed distribution. The model is specified as follows: $$PauseDuration \sim s(UtteranceLength, by=cohort) + UtteranceFinal$$ The model estimate of the average duration of pauses in non-final positions is 411 ms in English and 444 ms in Korean. Only older speakers of English seem to deviate from the average pause duration in non-final positions of longer utterances. In the model, the average pause duration in older speakers of English decreases linearly with utterance length (English, older: $edf=1, F=6.5620, p=.0104$, younger: $edf=1, F=0.031, p=.8610$, Korean: older: $edf=1, F=1.028, p=.311$, younger: $edf=1, F=0.429, p=.513$).

The Korean data does not support our hypothesis that experience ought to increase the burstiness of the pause distribution in older speakers. On one hand, this could be an effect of the relatively large differences in the age range of the English and Korean cohorts. On the other, as mentioned above, the temporal structure of Korean speech sequences differs fundamentally from the temporal structure of English. Alternation in syllable duration is an important functional feature of the sequential structure in English: the variability in stress patterns is a source of information. Korean, by contrast, is markedly less variable in vowel (and syllable) duration and appears to approach an isochronous temporal structure similar to mora-timing in Japanese \citep{tark2012experimental, moon2004rhythm}. This suggests that the functional load on alignment in timing might be substantially reduced and more evenly distributed across other functional features of the signal in Korean. To explore whether the differences in the cohort effects reflect a general difference in the distribution of pauses and durational contrasts in English and Korean, we conducted a final set of sampling simulations to address this question.

\subsection{Contrasting age-cohort differences in pause variability and pause duration in consecutive and random samples}

\begin{figure}[ht]

\centering
 \includegraphics[width=0.99\textwidth]{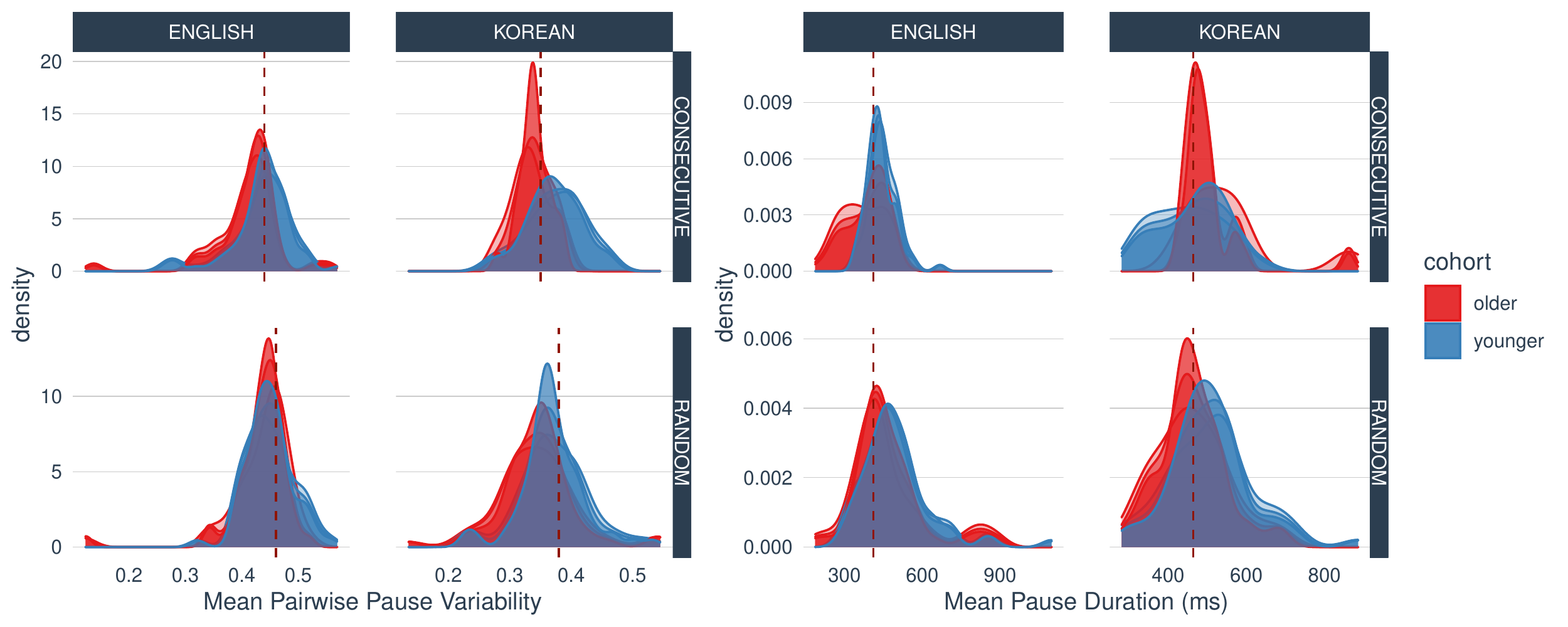}
 \caption{Distribution of mean nPVI (left panel) and mean pause duration (right panel) in blocks of consecutive (top row) or randomly sampled events (bottom row) for older (red) and younger (blue) speakers. The dashed red line marks the sample mean. Mean values from random samples converge on a normal distribution that centers on the sample mean. Means from samples of consecutive pauses suggest differences between the cohorts and the languages. In Korean, there is more dispersion in younger than in older speakers' consecutive samples, which could indicate that local and global patterns of pause distribution get more similar (less sparse/bursty) across the lifespan. By contrast, the distribution of averages from the English sample turns increasingly bimodal with speaker age and interview time in consecutive samples, indicating an increase in local bursts of shorter pauses in older speakers and longer pauses in younger speakers that also increases throughout the interview (shaded areas show the distribution of samples from later blocks).  The divergent patterns between the languages suggest that alignment is achieved through local optimization (bursts of activity) in English and globally (through an increasingly uniform distribution) in Korean.}
 \label{fig:burstiness_experience}

\end{figure}

To examine the distribution of sample averages we obtained means from blocks of consecutive pause durations and normalized durational contrasts (\textbf{nPVI}) from older and younger speakers' speech. We simultaneously extracted means from blocks of randomly ordered pauses to serve as a baseline. Figure~\ref{fig:burstiness_experience} shows the distributions of mean pause duration and mean nPVI for blocks of 20 pauses. Pause duration averages from random samples approximate a Gaussian distribution that centers around the theoretical mean in Korean and English. The mean nPVI in random samples is higher than the mean nPVI in consecutive pauses in English ($M_{ran}=0.4362, M_{cons}=0.4267, U=0.2564, p<0.01$). The difference between the model nPVI and the empirical nPVI indicates that the order in which pauses from different ranges of duration occur decreases the average distance between adjacent pauses in English, but not in Korean ($M_{ran}=0.3595, M_{cons}=0.3572, U=0.0815, p=.15$). 

In the cohort comparison, the distribution of nPVI averages from samples of consecutive pauses indicates local reductions of durational contrast in older speakers of English ($M_{o}=0.417, M_{y}=0.435, U=0.1958, p<.001$) and Korean ($M_{o}=0.34, M_{y}=0.378, U=0.2564, p<.001$). There is a significant difference between the random and the empirical distribution of pairwise variability in English ($M_{ran}=0.436, M_{cons}=0.427, U=0.147, p<.01$). We find no such difference in Korean ($M_{ran}=0.359, M_{cons}=0.357, U=0.081, p=.15$). This 'clumpiness' in the relationship between consecutive pauses in English suggests that the order in which temporal events unfold is more informative in English than in Korean. This is consistent with findings that reveal differences in the extent to which language users exhibit preferences in perceptual grouping of rhythmical sequences. English speakers are sensitive to the relationships between durations of consecutive events, whereas speakers of the more rhythmically consistent Japanese do not exhibit strong grouping preferences \citep{iversen2008perception}. In other words, temporal fluctuations of consecutive intervals constitute signals in English. This does not appear to be the case in Korean (or Japanese).

The distribution of pause durations from consecutive samples also reveals differences between the languages. In English, distributions of averages are bimodal (suggesting local bursts of activity) and show diverging patterns in older and younger speakers: bursts of shorter than average pauses in older speakers and longer than average pauses in younger speakers. This tendency increases in the later interview blocks, suggesting that the effect is cumulative and develops at multiple timescales (utterance, interview time, and speaker experience). In Korean, we find the opposite effect: older speakers' averages are markedly less dispersed and show more convergence across local samples, which can mean that the variability in pause duration decreases globally in Korean. Younger Korean speakers' samples appear to be far more variable in average duration and look more similar to older English speakers' distribution. The dispersion in consecutive samples we see in Figure \ref{fig:distributions} and table \ref{sample-table} appear to reflect bursty patterns produced by older speakers of English and younger speakers of Korean. We interpret these sample differences as divergent effects of lifelong experience on the distribution of pauses in spontaneous Korean and English. In English, differences between older and younger speakers increase at those points where experience-driven changes are greatest, at the boundaries of longer phrases. The results of the analyses indicate that patterns of pause production systematically interact with \textit{local} changes in information rates, providing support for our hypothesis. In Korean, contrary to our prediction, experience appears to lead to a \textbf{global} decrease in the temporal variability of pause durations. Given the many uncertainties that follow from the differences between the languages and the speech samples at our disposal (the corpora are similar in many ways, yet not strictly parallel), however, these findings cannot serve as strong evidence of functional interactions between pause and information rate. Instead, they can serve to inform future research.
    
\section{Summary and Discussion}

In this article, we examined the hypothesis that speech pauses play a crucial role in the systematic temporal structuring of speech signals and the maintenance of mutual predictability in time. We presented a theoretical rationale for the hypothesis and evidence to support it. We suggest that pauses provide information about fluctuations in articulation rates and that predictable interactions between pause duration and information rate facilitate alignment between speakers at different levels of experience. We approach the problem of alignment as a task of learning models that maintain mutual predictability of signals transmitted across multiple timescales. These models are learned from exposure to a noisy signal and structured by generations of speakers. We argue that a time-invariant source of information is necessary to counter the inevitable misalignment of expectations (i.e., differences in experience) that this complex distributed learning entails. We highlight how the distribution of information in human vocal signals contributes to learnability and efficient transmission across the lifespan.

We proposed memorylessness as a key prerequisite for successful alignment and efficient transmission and our finding that pause distributions from both speech sources, Korean and English, closely approximate the exponential distribution thus appears to fulfill this requirement. We hypothesized that to enable alignment in the way suggested, the exponential source ought to exhibit the following three properties: 

\begin{enumerate}
    \item a global distribution that provides developing speakers with a consistent source of information in the noisy signal they are exposed to
    \item a time-invariant source of information to ensure that all speakers will acquire and maintain sufficiently similar models of expectations independent of the points in time they enter the speaker community
    \item systematic local fluctuations in the temporal relationships between pauses to allow adult speakers at different levels of experience to rapidly adapt to local changes in segmentation rates
\end{enumerate}
  
Our results indicate that pauses meet all three requirements. Their empirical distributions increase the sampling efficiency in random samples, and their global distribution is characterized by local fluctuations, which lead to a skew in the distribution of sample averages from sequences of consecutive pauses. To investigate whether these local fluctuations are systematically correlated to variation in the distribution of information, we analyzed the distribution of pauses across sequences of different lengths, comparing speech samples of younger and older adults.

We found that experience increases the average utterance length in conversational speech and that this change leads to a systematic redistribution of information across utterance positions. We suggested that this redistribution of information reflects a decrease in uncertainty about structural regularities that set the utterance context and increase the efficiency of message transmission. The result is a difference in the distribution of information between the cohorts that increases linearly as a function of utterance position. The distribution of pauses, by contrast, does not change significantly with experience. The only change we observe in the distribution of pauses is a gradual reduction in the frequency of pauses that diverge notably from the mean pause duration indicating a global reduction of temporal contrast in pauses from the middle range of durations (250 - 750 ms) and a regression to the mean. Notably, we observe almost identical patterns of change in the cumulative distributions of English and Korean. Further, these effects are consistent with the predictions of the theoretical model derived from the quantitative structure of languages and learning theory.

Our results show that the misalignment in the distribution of information across utterances in relation to the relatively stationary distribution of pauses leads to a predictable decrease in the average pause duration in longer utterances produced by older speakers. This finding indicates a shift in the relationship between pause duration and the utterance position that is systematic -- it scales linearly with utterance position -- and is thus consistent with the suggestion that the relationship between pause duration and the changes in time intervals at which informative changes in the signal are observed are predictable.

Importantly, we find different patterns of age-related changes in the spread of Korean and English pause distributions. We hypothesized that the rates at which speakers segment the signal across sequences would become more variable with experience, increasing the signal sparsity (burstiness) over time. The effect is evident in the English sample, but not in the Korean sample. Contrary to our hypothesis, older speakers of Korean appear to minimize durational contrasts globally. By contrast, English speakers appear to minimize durational contrasts locally as predicted. Retrospectively, given that our hypothesis was that pause durations provide an alignment signal through consistent interaction with segmentation rate, the divergence in the patterns of lifelong development between the two languages need not be surprising. Our hypothesis was informed by evidence of the cumulative effect of experience on the performance of English speakers, consistent with the vocabulary development specific to the relatively impoverished information structure of English morpho-syntax \citep{ramscar2014myth}. However, Korean and English differ significantly in the extent to which patterns of lexical productivity are implicit (characterized by regularities such as a rich morphology) or explicit (characterized by less productive, explicitly lexicalized forms) \citep{ramscar2021discriminative}. Korean relies heavily on a rich morphological structure to maintain systematic variation in the distribution of word forms. English morphology, by contrast, is far less informative such that far more meanings are realized in explicitly lexicalized, often idiosyncratic forms. From a learning perspective, this has a differential impact on lifelong development of vocabularies and the distribution of functional load across speech sequences (contexts) in Korean and English.

One possible explanation of the effect is as follows: Language learning relies on regularities, such as for example, relative invariance in the patterns of inflection \citep{ramscar2013suffixing}.  Unattested forms, e.g., regular plurals, can often be easily inferred from the general pattern of inflection (squid-squids, octopus-octopuses, wug-wugs, niz-nizzes), providing language learners with an important source of predictable variance. By contrast, irregular forms can only be learned explicitly \citep[see][for review]{ramscar2021discriminative}. By implication, the fact that the relatively impoverished morphological structure of English relies more on word forms that cannot be derived implicitly leads to a redistribution of functional load to those aspects of the signal that support the learning of explicit forms. Conversely, in Korean the functional load ought to be distributed across a variety of morphological cues (particles, endings and functional prefixes). 

We have argued that the learning of explicit word forms and the alignment of speaker expectations in a communicative system characterized by non-linearity are facilitated by context. Because context is set both by the highly variable aspects of communicative codes such as regular patterns of co-occurrence between words, phrases, and segments \citep{ramscar2019source, linke2020grammar}, but also the less variable aspects of speech signals such as prosodic stress patterns and the variability in duration \citep{mcqueen2020prosody}, the differences in the lexical productivity of word forms we have described can result in very different predictions when it comes to the dynamics of lifelong learning. It is thus not entirely surprising that the experience-related changes in the patterns of pause distribution lead to increasingly uniform signal distributions in Korean and increasingly sparse signal distributions in English. We suggest that these differences in pause distributions could reflect differences in the dynamics of lifelong learning that interact with the linguistic structure set by the morpho-syntax and prosody/timing. These considerations highlight the necessity of extending these analyses to other languages that are placed along what seems to be a gradient scale of structural organization in human communicative codes. 

\section*{Data Accessibility}
The Buckeye and the Seoul corpus are freely accessible corpora for non-commercial uses. Extracted data used in this analysis and the analysis code associated with the current submission are available at \url{https://doi.org/10.5061/dryad.n5tb2rbxr.} Any updates will also be published in the Dryad Digital Repository. 

\section*{Disclosure of interest}
The authors report no conflict of interest.

\bibliography{main}

\end{document}